%% file: main.tex
\documentclass[runningheads]{llncs}
\usepackage[square,numbers]{natbib}
\usepackage{graphicx}
\usepackage{booktabs} % For formal tables
\usepackage{times}
\usepackage{helvet}
\usepackage{courier}
\usepackage{amsfonts}
\usepackage{graphicx}
\usepackage{amsmath}
\usepackage{amssymb}
\usepackage{subfigure}
\usepackage{color}
\usepackage{algorithm}
\usepackage{algorithmicx}
\usepackage{algpseudocode}
\usepackage{caption}
\allowdisplaybreaks

\begin{document}
\title{Path Planning Games}
\author{Yi Li \and  Yevgeniy Vorobeychik}
\institute{Electrical Engineering and Computer Science,
Vanderbilt University, Nashville, TN}%, Princeton NJ 08544, USA \and
%\email{@vanderbilt.edu}
%
\maketitle              % typeset the header of the contribution
\begin{abstract}
Path planning is a fundamental and extensively explored problem in
    robotic control.
We present a novel \emph{economic} perspective on path planning.
%there has been little research investigating the issue of multiple path planners facing a possible strategic conflict.
Specifically, we investigate strategic interactions among path planning agents using a game theoretic path planning framework.
    Our focus is on economic tension between two important objectives: efficiency in the agents' achieving their goals, and safety in navigating towards these.
    We begin by developing a novel mathematical formulation for path
    planning that trades off these objectives, when behavior of other
    agents is fixed.
 %   computing 
%a best response path for an agent given a fixed path plan of others in a stochastic environment.
    We then use this formulation for approximating Nash equilibria in path
    planning games, as well as to develop a multi-agent cooperative
    path planning formulation.
% through a best response dynamics algorithm.
%    Finally, we develop a novel multi-agent path planning formulation for computing a social welfare optimizing multi-agent path plans.
    Through several case studies, we show that in a path planning
    game, 
%tension between efficiency and safety can be resolved in a socially
%suboptimal way, with 
safety is often significantly compromised compared to a cooperative
solution.
%, even when all agents have a strong interest in avoiding collisions.

\end{abstract}
\input{intro}

\input{relatedwork}

\input{model}

\input{br}

\input{equilibria}

\input{optimal}

\input{experiments}
\small
\bibliographystyle{splncsnat}
\bibliography{mybibliography}

\end{document}

%% file: intro.tex
\section{Introduction}

Path planning is a fundamental technical problem in autonomous robotic control.
Decades of development have led to significant theoretical and algorithmic progress, with autonomous vehicles (including autonomous cars and UAVs) increasingly finding their way to urban roads and skies.
%Path planning, one of most significant challenges confronting autonomous agents, arouses the research interests in the area of artificial intelligence and autonomous control. Several decades of development has led to many theoretic foundations including both analysis techniques and algorithms. 

In much of the research on path planning, including mobile robot navigation \citep{DeSouza2002,Arkin1989,Elfes1989}, a fundamental task is to find a collision-free motion from a starting position to the goal position given a collection of known obstacles. 
Variations on this theme, such as dealing with stochastic and moving obstacles, have received recent attention with the emergence of numerous novel unmanned robotic systems and aerial vehicles~\citep{Mahony2012,AuatCheein2013,Craighead2007}.
%Recently, with the appearance of various novel unmanned robotics and aerial vehicles \cite{Mahony2012} \cite{AuatCheein2013}, path planning has evolved to address a large number of variations on the path finding problem. 
%\cite{Craighead2007}, path planning has evolved to address a huge number of variations on the path finding problem. 

As interactions among autonomous vehicles, be it on our roads or in the skies, becomes more routine, we can expect a certain amount of conflict to emerge, as the autonomous agents, designed in service of their individual goals, must occasionally find these goals dependent on other autonomous agents nearby.
However, remarkably little research has been devoted to the question of what autonomous vehicle ecosystem would thereby emerge, when many autonomous agents attempt to achieve their individual goals, but must necessarily interact with one another in doing so.

To investigate the consequences of such \emph{strategic} interactions among multiple path planners, we propose a study of \emph{path planning games}.
An important feature of such games is that a collection of self-interested path planners each trade off two objectives: efficiency, or speed with which their goals are achieved, and safety, or probability that they crash before reaching their goals.
Moreover, they trade these off in individual, potentially diverse, ways.
Consequently, in order to study path planning games we must take an \emph{economic}, rather than a purely algorithmic, perspective on path planning.

To this end, we first develop a novel mathematical programming method for computing a single-agent path plan, accounting for these two objectives, given fixed dynamic behavior (i.e., path plans) of all other agents, as well stochastic disturbances in the environment.
Next, we propose a simple iterative algorithm, best response dynamics, for approximately computing Nash equilibria of path planning games, given the best response mathematical programs.
Finally, we develop a novel mathematical program for computing a cooperative multi-agent path plan which optimally trades off efficiency and safety among all agents---that is, again, taking the economic perspective on the multi-agent path planning problem.

%We propose a new framework, path planning games, for solving multi-agent path planning problem. Three key contributions are made in this paper; 1) We introduce a conjunctive mixed integer linear programming solving the multi-agent path planning with motion uncertainty. 2) We address the trade-off between the agents' risk and performance by sparing the safety margin. 3) We numerically investigate the price of anarchy in path planning games with motion uncertainty.

We numerically investigate path planning games through several case studies involving two and three agents.
Our central observation is that as safety becomes more important to agents, a large gap opens up between safety achieved by a socially optimal and Nash equilibrium outcomes; in other words, Nash equilibria exhibit significantly more collisions than desirable by all agents.
The main reason for this is that while each agent is concerned with safety, they only account in their objective for the impact of collisions on themselves, and not on other agents who crash along with them.

Our observation about safety consequences of path planning games raises a concern as we look towards the future of autonomous vehicles interacting in populated environments, particularly as they tend to be designed primarily in service of their individual ends, rather than those of the entire autonomous and non-autonomous vehicle ecosystem.

%% file: relatedwork.tex
\section{Related Work}
%\paragraph{Multi-Agent Path Planning}
One common paradigm for studying multi-agent path planning problems is by considering cooperative path planning involving multiple agents.
For example, \citet{shen2008game} studied cooperative path planning in UAV control system, while \citet{lavalle2000robot} presented an algorithm for applying path planning with stochastic optimal control. %and multiplayer games. 

Game theoretic problems related to path planning have been considered from several perspectives.
Closest to traditional path planning are zero-sum models of games against nature in which agents are designed to be robust against adversarial uncertainty in the environment~\citep{chen2014path,chen2014multiplayer}.
Classic approaches consider rules of interaction and negotiation among self-interested agents, including planning agents~\citep{rosenschein1994rules,jonsson2011scaling,jordan2017better}.
Loosely related also is the extensive literature on multi-agent learning, in which multiple agents repeatedly interact in strategic scenarios in which rewards and dynamics depend on all agents (often modeled as stochastic games)~\citep{stone2000multiagent}.

Another important class of game theoretic models related to path planning are \emph{routing games}.
%The multi-agent routing problem is similar with path planning problem which is well studied. 
The routing games, as a framework for modeling routing traffic in a large communication network, were first informally discussed by \citet{Pigou1932}. This model was first formally defined by~\citet{wardrop1952road} based on a flow network under the non-atomicity assumption. 
Therefore, equilibrium flows in non-atomic selfish routing games are often called \emph{Wardrop equilibria}. Since then, a number of fundamental results for the non-atomic routing games have been proved by various researchers, such as the existence and uniqueness of equilibrium flows \citep{beckmann1956}, first-order conditions for convex programming problem \citep{Bertsekas1999}, and the theory of general non-cooperative non-atomic games \citep{Schmeidler1973}. 
The seminal work by \citet{Roughgarden2000} first characterized the gap between centralized and decentralized control in multi-agent routing problems, formalized as the price of anarchy, or ratio of socially optimal to worst-case equilibrium outcomes. Their work explained the principles behind a broad class of counter-intuitive phenomena, such as Braess's Paradox \citep{Braess1968}. 

Both routing games and path planning games investigate the competition among agents during their navigation tasks (e.g. passing through bottlenecks). However, in routing games, the state space is a graph-based structure, and the cost of competition is modeled by a set of latency functions without considering the agents' dynamics, while path planning games consider the problem at higher fidelity, with a continuous state space where the latency is caused by the interaction among agents.
Moreover, our model of path planning games allows us to explicitly study the tradeoff agents make between performance and safety, an issue not considered in routing games.

%% file: model.tex
\section{Model}
We describe the problem by first introducing the model of agents' motions, and then formulating the path planning game.

Consider a state space $\mathcal{X} = \mathbb{R}^n$. 
%We assume that $i$th agent is covered by a polyhedron represented by a set of $M_i$ hyperplanes, i.e., 
We represent an agent $i$ by a polyhedron  described by a collection of $M_i$ hyperplanes: $P_i = \{a_{ij}^Tx \le b_{ij}, j \in \{0,..., M_i\}\}$.
%constrained by $M_i$ hyper planes $a_{i,j}^Tx = b_{i,j}, j \in \{0,..., M_i\}$, and 
Each agent polyhedron $P_i$ contains 
%has 
a point $r_i\in\mathcal{X}$ called the \emph{reference} which rigidly attaches to the polyhedron such that the state of an agent can be determined by the position of its \emph{reference}.
We assume that agents move in discrete time, 
%Assume agents' motions happen under a discrete time horizon 
and a control input $u_{it} \in \mathcal{U}_{it} \subset \mathbb{R}^m$ applied to the $i$th agent at time $t$ moves the agent from state $r_{i,t}\in\mathcal{X}$ at time $t$ to state $r_{i,t+1} \in \mathcal{X}$ at time $t+1$ according to a linear stochastic dynamic model
%\begin{equation*} 
%r_{i,t + 1} = f(r_{i,t}, u_{i, t}, \omega_{i}).
%\end{equation*}
%Here we further assume that $f$ is a linear function,
\begin{equation}\label{motion_dyn}
%f(r_{i,t}, u_{i, t}, \omega_{i}) 
r_{i,t + 1} = A_{i}r_{it} + B_{i}u_{it} + \omega_{i},
\end{equation}
where $A_{i} \in \mathbb{R}^{n\times n}, B_{i} \in \mathbb{R}^{n \times m}$, and $\omega_{i} \sim \mathcal{N}(0, \Sigma_{i})$ is the process noise for $i$th agent at time $t$ following an $n$-dimension zero-mean Gaussian distribution with a covariance matrix $\Sigma_{i}$.

For each agent we are given its initial placement $r_0 \in \mathcal{X}$ (i.e., where the agent starts) and a goal $r_{goal} \in \mathcal{X}$ which the agent needs to reach. 
Let $r_{i,0:T} = <r_{i0},...,r_{iT}>$ be a state sequence of the (reference point of the) $i$th agent from time 0 to $T$ and $u_{i,0:T} = <u_{i0},...,u_{iT}>$ be a corresponding control sequence.
%control sequence of i-th agent from stage 0 to $T$. 
However, once the agent reaches its goal, it remains there deterministically, and has no effect on other agents.
%In this stochastic motion model, since the future position of agents at a given time step is a random variable,
We aim to find the optimal control sequence for the $i$th agent in this stochastic motion model, with the following criteria in mind:
%based on the following criteria:
\begin{enumerate}
    \item After applying the resulting control sequence, the expected terminal position of the $i$th agent is $r_{i,goal}$,
%Furthermore, it will not affect the others once its expected position reach to its goal position.
    \item the upper bound of the probability that the $i$th agent collides with other agents should be minimized, and
%considered and the smaller is good (risk concern).
    \item the agent reaches the goal in as few time steps as possible.
%Simultaneously, the resulting control sequence should lead the $i$-th agent to reach its goal as fast as possible (performance concern).
\end{enumerate}
For the moment, we allow no feedback from observed state to control; we relax this restriction below.

\noindent{\bf Path Planning Game:} 
Given these models of individual agents, we define a \emph{path planning game} by a collection of $N$ agents, with each agent $i$'s action space 
%where each agent $i$ has the action space 
comprised of all possible control sequences, $\prod\limits_{t=0}^{T}\mathcal{U}_{it}$.
In this game, each agent aims to compute an optimal control sequence, given the behavior of others, trading off two objectives: efficiency, or the number of times steps it takes to reach the goal, and safety, or the probability of collision.
To formalize, let $T_i$ be the expected number of times steps to reach the goal (if no collision occurs), and $G_i$ the \emph{safety margin}, related to the upper bound on the probability of collision as discussed below.
%or the upper bound on the probability of collision.
An agent $i$'s objective is then
\begin{equation}
J_i(u_{i,0:T_{max}}, u_{-i,0:T_{max}}) = \lambda T_i + (1-\lambda) G_i,
\end{equation}
%Below we discuss in detail how these entities are computed, and how an agent then solves a single-agent optimization problem.
What makes this a game is that the safety $G_i$ of an agent $i$ depends on the paths taken by \emph{all agents}, rather than $i$ alone.
For example, if two agents are moving towards one another, and directly towards their respective goals, the only way for one of them to avoid collision is to circumnavigate the other, taking a longer path towards the goal. 
Next, we describe how to define and compute $T_i$ and $G_i$, and compute a \emph{best response} for a given agent $i$, fixing behavior of all others.

%Hence, such multi-agent system can be treated as a $N$ players game. For the $i$-th player, its action space consists of all possible control sequences, which is $\prod\limits_{t=0}^{T_{max}}\mathcal{U}_{i,t}$ and $T_{max}$ is the maximum time horizon. By considering the performance metric as the time step consumed by the agent to reach to its goal position and safety metric as the upper bound of possibility of collision in each time step, the utility function that the $i$-th player want to minimize can be defined as
%\begin{equation}
%J_i(u_{i,0:T_{max}}, u_{-i,0:T_{max}}) = \lambda T_i + (1-\lambda) G_i
%\end{equation}
%where $T_i$ is the consumed time step and 
%\begin{equation}
%Pr(collision) < G_i
%\end{equation}
%in each time step.

%% file: br.tex
\section{Computing an Agent's Best Response}

An important subproblem of computing a Nash equilibrium of a path planning game is to compute a best response of an arbitrary agent $i$ when we fix the control policies of all others.
%In the path planning game, the best response of the current agent is the optimal control sequence by given the control sequences of the others, which is a sub-problem of finding the equilibrium in path planning games via the best response dynamic process. 
We show that calculating agents' best responses in path planning games
amounts to a single-agent path planning problem with motion
uncertainty.
%, a novel variant on stochastic path planning.
%which is a novel variant of well-studied problems modeled by stochastic system. 
%By considering probabilistic obstacles, we can convert this problem
%into a deterministic system by shrinking the feasible planning domain.
% (through extending the margin of obstacle). 
%Based on the probabilistic obstacle approach, such stochastic system can be further turned to a approximate deterministic system by shrinking the feasible planning domain (extending the margin of obstacle). 
%By leveraging this idea, 
\citet{blackmore2006probabilistic} previously
developed a probabilistic approach for computing a robust optimal path for a robot in the environment with a static obstacle and motion uncertainty via mathematical programming. 
However, in our context, where an agent trades off efficiency and
safety, with stochastic \emph{moving} obstacles (representing other
agents), this prior approach is inadequate.
%In this section, we further extend \cite{blackmore2006probabilistic}'s work to fit our problem.
In this section we develop a novel method for solving such problems.
%,
%building on the ideas by Blackmore et al.~\cite{blackmore2006probabilistic}.
%from \citeauthor{blackmore2006probabilistic}~\shortcite{blackmore2006probabilistic}.

\subsection{Best Response for a Point-Like Agent}
\begin{figure} [t!]
  \centering 
  \subfigure[]{ 
    \label{fig:subfig:a} %% label for first subfigure 
    \includegraphics[width=0.22\textwidth]{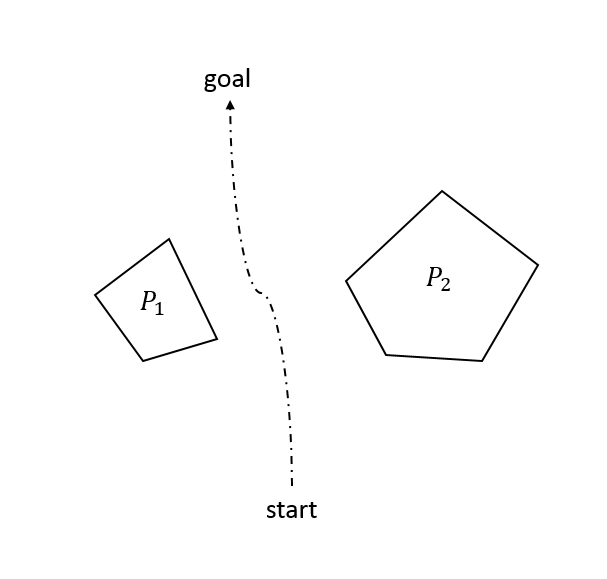}} 
  \hspace{1mm} 
  \subfigure[]{ 
    \label{fig:subfig:b} %% label for second subfigure 
    \includegraphics[width=0.22\textwidth]{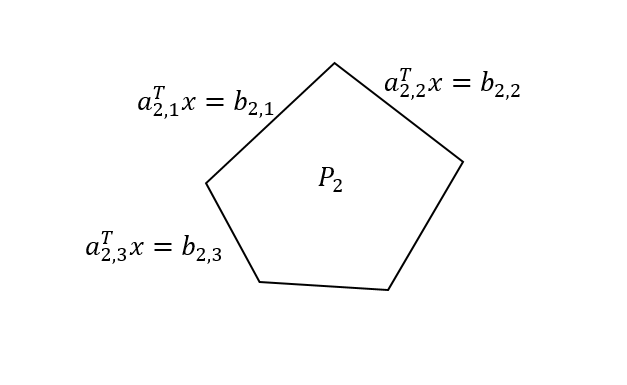}} 
   %% label for entire figure 
  \caption{Single agent path planning with a point-like agent.}
  \label{fig:sp11}
\end{figure}
First, consider a simple path planning problem illustrated in Figure \ref{fig:sp11}. In this problem, there is a set of static obstacles and an agent, represented by a point, aiming to find a collision-free minimum-time path from its initial placement to its goal position under motion uncertainty. 
Assume each obstacle has a given collision volume which can be represented by a polyhedron. 
To create a mathematical program for solving this problem, two factors need to be taken into account: goal position constraints and collision avoidance constraints. 

Formally, let $r_t$ denote the position of an agent at time $t$ with its initial placement $r_0$ and the goal position $r_{goal}$.
Suppose that the motion dynamics of the agent follows
(\ref{motion_dyn}) (from which, we remove the index $i$, since there
is only one agent). Assume there are $K$ obstacles represented by polyhedra $P_n, n = 1,...,K$, with $P_n = \{x|a^T_{np}x \le b_{np}, p =1,\ldots,E_n\}$, where $E_n$ is the number of hyperplanes representing the polyhedron $P_n$.
%each of which constrained by $E_n$ hyperplanes, denoted by $a^T_{n,p}x = b_{n,p}, p = 1,...,E_n$. 
As before, let $T$ denote the planning horizon (so that the goal must be reached by time $T$; we assume the horizon is long enough that the goal can be successfully reached even with the obstacles).
%We use $T_{max}$ to represent the maximum time horizon. 
%Since the programming process is supposed to return a finite control
%sequence $u_{0:T}$, we can assume $T_{max}$ is always large enough to
%have no effect to the solution.\\

\noindent{\bf Efficiency and Reachability:}
%\textbf{goal position constrains:}\\
Let $\{d_{0},...,d_T\}$ denote a collection of binary indicators which indicate whether the agent has reached its goal, i.e., $d_t = 1$ iff $r_t = r_{goal}$.
% is on its goal or not and
%$$d_{t} = \left\{
%\begin{array}{ll}
%    0 & \quad  r_{t} \neq r_{goal}\\
%    1 &  \quad  r_{t} = r_{goal}
%    \end{array}
%    \right.
%$$
Then, with a large positive number $M$, the constraints
\begin{align}
\forall t, &||r_t - r_{goal}|| \leq M(1-d_t)\label{E:posConstraint}\\
&\sum\limits_{t = 0}^{T} d_t = 1
\end{align}
make sure that the agent will reach to its goal position sooner or later (and we assume that there exists a feasible solution). 
Moreover, the number of time steps to reach its goal position can be represented by
\begin{equation}\label{eq:objective}
T = \sum_{t = 0}^{T} t \cdot d_t
\end{equation}
%which is the value we want to minimize.
which is one of our objectives (corresponding to $T_i$, for an agent $i$ above). 
Since $r_t$ is a random variable, this constrain is stochastic. We approximate it by a deterministic constraint, replacing the position of the agent $r_t$ with its expected position $\overline{r}_t$ in Constraint~\eqref{E:posConstraint}.
%\textbf{collision avoidance constrains:}\\

\noindent{\bf Collision Avoidance:}
Let $A$ denote the event that the agent has a collision, and let $A(n, t), n \in \{1,...,K\}$ denote the event that the agent collides with the $n$th obstacle at time step $t$. 
We wish to minimize the probability of a collision, $Pr(A)$, or minimize $G$ such that
\begin{equation} \label{eq:overall_risk}
Pr(A) \leq G.
\end{equation}
%$G$ is the value of risk which we want to minimize. 
%We say that the 
The agent has a collision if the agent collides with any of obstacles
at any time steps, which is the event
%we capture as
\begin{equation}
A = \bigvee\limits_{t= 0}^{T}\bigvee\limits_{n = 1}^K A(n,t)
\end{equation}
Then, by the union bound
%\begin{equation*}
%P(A \vee B) \le P(A) + P(B)
%\end{equation*}
%we obtain
\begin{align}
    &Pr(A) = \Pr\left(\bigvee\limits_{t = 0}^{T}\bigvee\limits_{n = 1}^K A(n, t)\right) \leq \sum\limits_{t = 0}^{T} \sum\limits_{n=1}^K \Pr(A(n, t)) \leq G\\
    &\Leftarrow [\forall n,t,Pr(A(n, t)) \leq g(n, t)] \wedge [\sum\limits_{t=0}^{T} \sum\limits_{n=1}^K g(n, t) = G],
\end{align}
%We use $g = \frac{G}{K}$. 
where $g(\cdot)$ is \emph{risk allocation} which indicates how the risks are distributed among obstacles and time steps.
%we can minimize $G$ by minimizing $g$.
Next, we consider the event that the agent collides with an obstacle at time step $t$, which means that the position of the agent is inside the corresponding polyhedron. 
%The the agent collides with the $n$-th obstacle surrounded by $E_n$ number of hyper planes at time step $t$ can be describe by 
Thus, collision with the $n$th obstacle can be described by 
\begin{equation} \label{eq:collision}
A(n, t) : \bigwedge\limits_{p = 1}^{E_n} a^T_{np} \cdot r_t \leq b_{np}
\end{equation}
Since the condition (\ref{eq:collision}) including $r_t$ is also stochastic, to convert it into a deterministic one, we consider its probabilistic measure, $\Pr\{A(n, t)\}$. 
Following (\ref{eq:overall_risk}), our constraints then become
%if we want the probability of collision with an obstacle for a time steps less equal than $g$, we get
\begin{equation}\label{eq:risk}
\Pr\left\{\bigwedge\limits_{p = 1}^{E_n} a^T_{np} \cdot r_t \leq b_{np}\right\} \leq g(n, t).
\end{equation}
Since a polyhedron is convex, a sufficient condition is,
\begin{equation}\label{eq:risk2}
\bigvee \limits_{p = 1}^{E_n} \Pr( a^T_{np} \cdot r_{t}\leq b_{np}) \leq g(n, t).
\end{equation}
%recall that now $g$ is the value of risk that we want minimize.
Based on the approach by \citet{blackmore2006probabilistic}, expression~\eqref{eq:risk} can be further simplified using the linear approximation of the upper bound on the probability of collision. 
First, consider $r_t$, the position of agent at time step $t$ given its initial placement $r_0$ and the control sequence $u_{0:t}$, which is a random variable following a Gaussian distribution, $r_t \sim N(\overline{r}_t, \Sigma_t)$, where
\begin{equation}
\overline{r}_t = \sum\limits^{t-1}_{k=0} A^{t-k-1}Bu_{k} + A^tr_{0}
\end{equation}
and
\begin{equation}
\Sigma_{t} = \sum\limits^{t-1}_{k=0}A^{t-k-1}\Sigma(A^T)^{t-k-1}.
\end{equation}
For a single Gaussian random variable $X \sim N(\mu, \sigma^2)$, we can take the inverse Gaussian distribution function at both sides of $\Pr(X < 0) \leq \delta$ and get $u \geq \sqrt{2}\sigma erf^{-1}(1-2\delta)$. 
Similarly, from $r_t \sim N(\overline{r}_t, \Sigma_t)$, we can get $(a^T_{np}r_t - b_{np}) \sim N(a^T_{np}\overline{r}_t - b_{np}, a_{np}^T\Sigma_ta_{np})$. 
Then, we take the inverse Gaussian distribution function at both sides of (\ref{eq:risk2}), and
\begin{equation}\label{eq:safety margin}
\bigvee \limits_{p = 1}^{E_n} a^T_{np} r_{t} - b_{np} \geq e(n, t)
\end{equation}
where $e(n, t) = \sqrt{2a^T_{np} \Sigma_{t} a_{np}} \cdot \mathit{erf}^{-1}(1 - 2g(n, t))$ and $\mathit{erf} (z) = \frac{2}{\sqrt{\pi}} \int^z_0 e^{-t^2} dt$.
We call this the \emph{safety margin}, because it expands the margin
of obstacles and shrinks the feasible planning domain in order to
consider motion uncertainty. 
Because the motion of the agent after it reaches its goal has no further
effect, we add the term $M \sum\limits_{k=0}^t d_k$ to these constraints where $M$ is a large positive number. 

Define $s(n, t) = \mathit{erf}^{-1}(1-2g(n, t))$.
Since $\mathit{erf}^{-1}$ is strictly monotonically increasing, we can minimize $\sum\limits_{t=0}^{T} \sum\limits_{n=1}^K g(n, t)$ by minimizing
\begin{equation}
G = -\sum\limits_{t=0}^T \sum\limits_{n=1}^K s(n, t).
\end{equation}
This is the \emph{safety} portion of an agent's objective ($G_i$ for
an agent $i$ above).

\noindent{\bf A Path Planning Mathematical Program:}
%We have two objective to minimize: $T$ and $G$. 
%If we make a compromise between $J_{time}$ and $J_{risk}$, according
%to both goal position constrains and collision avoidance constrains,
%the problem discussed in this subsection can be solved by following
%mathematical programming,
Our goal is to minimize $J = \lambda T + (1-\alpha) G$, balancing
efficiency and safety using an exogenously specified parameter $\lambda$.
%The actual objective of an agent weighs these two considerations with a parameter $\lambda$, with the goal of minimizing $J = \lambda J_{time} + (1-\lambda)J_{safety}$.
Combining this objective with the goal and collision avoidance constraints described above, we obtain the following mathematical program for single-agent path planning:

{\bf MP1:}
\begin{align}
&\min\limits_{u,s,d} \lambda T(d) + (1-\lambda) G(s)\\
\nonumber&\text{s.t.}\\
&\forall t, u_{t} \in \mathcal{U}_t\\ 
&\forall t, \overline{r}_{t} = \sum\limits^{t-1}_{k=0} A^{t-k-1}Bu_{k} + A^tr_{0}\\
&\forall t, ||\overline{r}_{t} - r_{goal}||_1 \leq M\cdot (1 - d_{t})\\
&\forall t, d_{t} \in \{0,1\}\\
&\sum\limits_{t = 0}^{T} d_{t} = 1\\
&\forall t\forall n, \bigvee\limits_{p = 1}^{E_{n}} a^T_{n,p} \overline{r}_{t}> b_{np}  + e(n, t) - M \sum\limits_{k=0}^t d_k \label{C:collision}\\
&\forall t, e(n, t) = s(n, t)\sqrt{a^T_{np}\Sigma_{t}a_{np}}\\
&\forall t, \Sigma_{t} = \sum\limits^{t-1}_{k=0}A^{t-k-1}\Sigma(A^T)^{t-k-1} \label{mp1:noise}\\
&\forall t \forall n, 0 \leq s(n,t) \leq M' \label{s:bound}
\end{align}
%where the output are, the control sequence $u_{0:T}$ and safety margin $s$. 
One residual concern is that if an agent cannot possibly collide with
an $n$th obstacle at time step $t$ (i.e., if $g(n,t) = 0$), $s(n,t)$
can become unbounded.
%has no chance to collide with $n$th obstacle at
%time step $t$ ($g(n,t) = 0$), theoretically, the value of $s(n,t)$ can
%be infinity (For example, whenever $\sum\limits_{k=0}^t d_k = 1$). 
To address this, we add Constraint~\eqref{s:bound} which imposes
an upper bound $M'$ on $s(\cdot)$, where $M'$ is an appropriate
positive number so that $\mathit{erf}(M') \simeq 1$. 

Since MP1 is a disjunctive linear program which can be solved by an
off-the-shelf linear programming solver.
%Since we only consider the
%control sequence leading the agent to its goal position, 
A solution
$<u,s(\cdot),d>$ found by MP1 with $d_{T_0} = 1$ means that the agent
can reach to its goal position in $T_0$ time steps with the
probability of collision at most $\sum\limits_{t=0}^{T_0}
\sum\limits_{n=1}^K \frac{1 - erf(s(n,t))}{2}$ by applying the control
sequence $u_{0:T_0}$. 

\subsection{Generalization: Feedback Control}

Above we considered \emph{open loop} path planning where the control
sequence is deterministic and fixed a priori.
We now extend our approach to \emph{closed loop (feedback)} control,
following the ideas in~\citet{geibel2005risk} and~\citet{oldewurtel2008tractable}.
%as an open loop control problem, where the yielded control sequence is deterministic, the positions of the agent are random variables and the uncertainty of its position grows over time since the system has no capability to fix the runtime deviation. To eliminate the accumulation of uncertainty, a close loop control approach is employed based on the work in \cite{geibel2005risk} and \cite{oldewurtel2008tractable}.

Assume we have a nominal control sequence $\overline{u}_{0:T}$. Then, the feedback control sequence can be obtained by integrating the nominal control sequence and the feedback gain:
\begin{equation}\label{eq:fbc}
u_{t} = \overline{u}_{t} + K(x_t - \overline{x}_t),
\end{equation} 
where $x_t$ is the observed and $\overline{x}_t$ the predicted
position, and $K$ is an exogenous parameter which determines the importance of
the error feedback term $(x_t - \overline{x}_t)$.
In this approach, $\overline{u}_t$ is computed using the MP1 offline,
and the actual control sequence is then generated at runtime by
applying~\eqref{eq:fbc}. 
As a consequence, the Constraints~\eqref{mp1:noise} above become 
%Then, after using ~\eqref{eq:fbc}, ~\eqref{mp1:noise} in MP1 will become
\begin{equation}
\Sigma_t = \sum\limits_{k=0}^{t-1} (A + BK)^{t-k-1} \Sigma [(A+BK)^T]^{t-k-1}.
\end{equation}
Notice that when there is no error feedback ($K=0$) this becomes
equivalent to open loop control.
%the feedback gain, $K$, equals to $0$), it is just the case with open loop control and the noise $\Sigma_t$ grows continuously over time. In the contrast, $\Sigma_t$ converges at some point when the largest eigenvalue of $(A+ BK)$ is less than $1$.

\subsection{Collision Avoidance for Polyhedral Agents}

%We now move on to the more complicated problem of dealing with an
%agent who is represented by a polyhedron, rather than a point.
Having considered the problem for point-like agents, and then
generalizing the approach to consider error feedback, we now
generalize the collision avoidance constraints to polyhedral agents.
%In this subsection, we discuss the cases where the agent is not point-liked (the agent has certain collision volume). 
%We show that we can generalize the collision avoidance constraints above.
%the collision avoidance constrains are still usable in these cases if we can "migrate" the collision volume of the agent to obstacles via $\mathcal{C}$-obstacle approach.

Consider states of the agent and the $n$th obstacle, both represented
by polyhedra $P_{t}$ and $P_{n}$, respectively. 
The position of the agents' reference is $r_{t}$. 
Since the reference point rigidly attaches to the agent, let $C = \{ x - r_{t}|x \in P_{t}\}$ denote the relative region of the agent to its time-dependent reference. 
When the agent collides with the $n$th obstacle at time $t$, we know
that $\exists x \in P_{t} \cap P_{n}$ (i.e., the intersection of these
time-dependent polyhedra is non-empty).
Thus, from the point view of the agent, the set of positions of its reference causing collision with the $n$th obstacle can be represented by
%\begin{align*}
\(K_{n} = \{x - c| x \in P_{n}, c \in C\}= -C \oplus P_{n},\) 
%\end{align*}
where $\oplus$ is the Minkowski addition. 
Since both $C$ and $P_{n}$ are polyhedra, $K_{n}$ is a polyhedron and can be represented by a set of hyperplanes:
%its margin is a set of hyper planes denoted by 
%\begin{equation} \label{Kmargin}
\(K_n = \{x | a^T_{np}x \le b_{np}, p =0,...,E_{n}\},\) 
%\end{equation}
where $E_{n}$ the number of hyperplanes of $K_{n}$. 
The agent collides with the $n$th obstacle at time step $t$ if the position of its reference is in $K_{n}$, that is, when
\begin{equation}\label{eq:non-point-liked agent}
r_{t} \in K_{n} \Leftrightarrow \bigwedge\limits_{p=1}^{E_{n}}a^T_{np}r_{t} \leq b_{np}.
\end{equation}
Comparing (\ref{eq:non-point-liked agent}) with (\ref{eq:collision}), we can see that the problem with polyhedral agents can also be solved via the mathematical program above, if we treat the agent as its reference point, and assign the collision volume $K_n$ to each obstacle. 

\subsection{Best Response Solver}
%In this subsection, we further expend the capability of MP1 to solve the agent's best response in path planning games by treating the agents with known control sequences as moving obstacles.

Our final challenge is to consider the actual best response problem of an arbitrary agent in the path planning game, where all other agents are \emph{moving} (rather than static) obstacles with known stochastic motion policies.
We now address this problem, obtaining the final mathematical program for computing a single-agent best response.

Let $i$ denote the agent for whom we are computing a best response,
with $-i = \{1,...,i-1,i+1,...,N\}$ the set of all others. 
%In the rest part of this section, we use $j$ denotes one of the agents which is other than the current one, such that $j \in -I$.
%Assume the current agent is the $i$-th agent, and $-i$ denotes every
%agent which is other than the current one. 
Let $i$ be represented by a
polyhedron $P_{it}$ with reference $r_{it}$ and let $j \in -i$ be
represented by $P_{jt}$ with reference $r_{jt}$.
Let $C_{i}$ denotes the relative region of $i$ to its reference, while
$C_{j}$ denotes the relative region of $j \in -i$ to its reference.
%their respective relative regions to their references. 
Suppose that $j$ reaches its goal position by time step $T_{j}$ with the corresponding known control sequences $u_{j,0:T_{j}}$. 
%The current agent is represented by 
Then, for each $j$ and $t$, $K_{ijt} = -C_i \oplus P_{jt}$ is a polyhedron with $K_{ijt} = \{x|a^T_{ijp} x \le b_{ijtp}, p \in \{0,...,E_{ij}\}\}$
%constrained by a set of hyper planes $\{a^T_{i,-i,p} x = b_{i, -i, t, p}\}, p \in \{0,...,E_{i,-i}\}$,
where $E_{ij}$ is the number of hyperplanes related to the shapes of $C_i$ and $C_{j}$. 

Now we formalize how the control sequence $u_{j,0:T_{j}}$ of each agent $j$ affects
$P_{jt}$ so that we can determine $K_{ijt}$. 
From motion dynamics of $i$ and $j$, 
\begin{align}
r_{it} &= \sum\limits_{k=0}^{t-1} A_i^{t-k-1}B_iu_{ik} + A_i^tr_{i0}+ \omega_{it}\label{eq:motion_current}\\ 
\forall j,\ r_{jt} &= \sum\limits_{k=0}^{t-1} A_{j}^{t-k-1}B_{j}u_{jk} + A_{j}^tr_{j0} + \omega_{jt} \label{eq:motion_others}
\end{align}
From the perspective of agent $i$, the motion of agent $j$ can be
treated as deterministic if we ``migrate'' motion uncertainty from $j$ to $i$ so that 
%, if the motion uncertainty in (\ref{eq:motion_others}) can be "migrate" to (\ref{eq:motion_current}) such that
\begin{align}
\forall j, r'_{ijt} &= \sum\limits_{k=0}^{t-1} A_i^{t-k-1}B_iu_{ik} + A_i^tr_{i0}+ \omega_{it} - \omega_{jt}\nonumber\\
\forall j, r'_{jt} &= \sum\limits_{k=0}^{t-1} A_{j}^{t-k-1}B_{j}u_{jk} + A_{j}^tr_{j0}.  \label{eq:motion_others2}
\end{align}
For each $j$, let $\omega'_{ijt} = (\omega_{it} - \omega_{jt}) \sim N(0,
\Sigma_{it} + \Sigma_{jt})$ denote the relative motion uncertainty of
$i$ to $j$ at time $t$.  
Let
\begin{align}
\forall j, \Delta r'_{jt} &= \sum\limits_{k=0}^{t-1} A_{j}^{t-k-1}B_{j}u_{jk} + A_{j}^tr_{j0}  -r_{j0}\label{eq:motion_others3}
\end{align}
denote the position shift of agent $j$ at time step $t$
determined by its control sequence $u_{j,0:T_{j}}$. Then, we obtain
the position of $K_{ijt}$ by shifting $K_{ij0}$ by $\Delta
r_{jt}$. 
Since 
%$K_{i,j,t}$ is captured by $E_{i,j}$ hyper planes,
%i.e., 
$K_{ijt} = \{x|a^T_{ijp} x \le b_{ijtp}\}$, we obtain
\begin{equation}
b_{ijtp} = b_{ij0p} + a^T_{ijp}\cdot\Delta r'_{jt}.
\end{equation}
%Finally, we can use MP2 to compute the best response of agent $i$ based on similar ideas in MP1, since
%\begin{itemize}
%\item  $i$ can be treated as a point-like agent with its own reference
%  point while the agents other than $i$ can be treated as moving obstacles with collision region $\forall j, K_{i,j,t}$.
%\item the position $K_{i,j,t}$ for each $j$ can be uniquely determined if the current agent follows the motion dynamic (\ref{eq:motion_others2}).
%\item the relative motion dynamics of $i$ to the others $j$,
%  $r'_{i,j,t}$, have the same expected motion dynamic, which means
%  $\forall j, \overline{r}'_{i,j,t} \equiv \overline{r}_{i,t}$.
%\end{itemize}
Consequently, we obtain the following mathematical program for $i$'s
best response:

{\bf MP2:}
\begin{align}
&\min\limits_{u,s_i(\cdot),d} J_i =  \lambda T_i+ (1-\lambda)G_i\\
\nonumber&\text{s.t.}\\
&\forall t, u_{it} \in \mathcal{U}_{it}\\ 
&\forall t, \overline{r}_{it} = \sum\limits^{t-1}_{k=0} A_i^{t-k-1}B_iu_{ik} + A_i^tr_{i0}\\
&\forall t, ||\overline{r}_{it} - r_{i,goal}||_1 \leq M\cdot (1 - d_{it})\\
&\forall t, d_{it} \in \{0,1\}\\
&\sum\limits_{t = 0}^{T} d_{it} = 1\\
&\forall j\forall t = 0,...,T_{j},\nonumber\\
&\bigvee\limits_{p = 1}^{E_{ij}} a^T_{i,j,p} \overline{r}_{it}> b_{ij0p} + a_{ijp} \cdot \Delta r'_{jt} + e_{ijt}\nonumber\\
&- M\sum\limits_{k =0}^t d_{ik}\label{mp2:others_motion}\\
&\forall i \forall t, \Sigma_{it} = \sum\limits^{t-1}_{k=0}(A_i + K_iB_i)^{t-k-1}\Sigma_i[(A_i + K_iB_i)^T]^{t-k-1} \label{mp2:noise}\\
&\forall t \forall j, \Delta r'_{jt} =  \sum\limits_{k=0}^{t-1} A_{j}^{t-k-1}B_{j}u_{jk} + A_{j}^tr_{j0} - r_{j0}\\
&\forall j, e_{ijt} = \sqrt{a^T_{ijp}(\Sigma_{it} + \Sigma_{jt})a_{ijp}}\cdot s_i(j, t)\\
&\forall t \forall n, 0 \leq s_i(n,t) \leq M'
\end{align}
Notice that the constraints (\ref{mp2:others_motion}) are effective only for
$t = 0,...,T_i$, and $i$ is not affected by any $j$ who reached its goal.
% which means the current agent will not be affected by those agents which have reached to its goal position. 

%% file: equilibria.tex
\section{Finding Equilibria in Path Planning Games}

%In this paper, we investigate agents' behaviors of multi-agent path planning problems in both non-cooperative and cooperative cases. In cooperative cases, we try to find the global optimum that minimizing the sum of all agents' utilities through a global path planner. In non-cooperative cases, we consider the situation where each agent knows the control sequences of the others and no one can gain further by changing its own one. Such solution concept is described by the Nash equilibrium of the path planning game.

Armed with the best response solvers for each agent $i$ in a path
planning game, our goal is to approximate a Nash equilibrium in the
resulting game.
We do so by applying \emph{best response dynamics} which, if it
converges (which it does in our experiments), yields a Nash equilibrium.

%The algorithm is formally shown as Algorithm~\ref{alg:BRD}.
Best response dynamics is an asynchronous iterative algorithm in which a single agent $i$ is
chosen in each iteration, and we maximize $i$'s utility (i.e., compute
its best response) fixing control strategies for all other agents.
Best response of an agent $i$ can be calculated as discussed above.
%Then, Nash equilibrium can be yielded by the best response dynamic (algorithm \ref{alg:BRD}),
%\begin{algorithm}
%\caption{Best Response Dynamics}
%\label{alg:BRD}
%\begin{algorithmic}[1]
%\State \textbf{input} initial action, $u^{(0)}_{1,0:T},...,u^{(0)}_{N,0:T}$
%\State \textbf{output} Nash equilibrium $u^*_{1,0:T},...,u^*_{N,0:T}$
%\State Set $k \leftarrow 0$
%\Repeat
%\For{\texttt{$i \leftarrow 1,...,N$}}
%\State $u^{(k+1)}_{i,0:T} = \arg\min\limits_{u_{i,0:T}} J_i(u_{i,0:T}, u^{(k)}_{N,0:T})$
%\EndFor
%\State Set $k \leftarrow k+1$
%\Until{$\forall i, u^{(k)}_{i,0:T} = u^{(k-1)}_{i,0:T}$}\\
%\Return {$u^{(k)}_{i,0:T}$}
%\end{algorithmic}
%\end{algorithm} 

%% file: optimal.tex
\section{Optimal Multi-Agent Path Planning}

We now extend the single-agent best response problem to compute an optimal multi-agent path plan.
%We compute the socially optimal path plans for multiple agents
%simultaneously via following mathematical programming.
%The optimal cooperative multi-agent path plan can then be computed
%using the following mathematical program.
In this case, the control sequences $u_{i,0:T_i}$ of all agents are
unknown a priori (as they are being computed jointly).
% and $u_{i,t}$ are undetermined variable. 
Compared to calculating an agents' best response, we replace the
objective of the current agent with the sum of all agents' objectives,
i.e., the new objective is $J = \sum_i J_i$, where $J_i$ is the
objective of agent $i$.
Moreover, 
%, and 
we add constraints analogous to MP2 to make sure that the collision
avoidance conditions hold from the perspective of \emph{every agent
  simultaneously}. 
We thus obtain the following mathematical program:

{\bf MP3:}
\begin{align}
&\min\limits_{u,s,d} J = \sum\limits_{i=1}^N J_i\\
\nonumber&\text{s.t.}\\
&\forall i, t, u_{it} \in \mathcal{U}_{it}\\ 
&\forall i, t, \overline{r}_{it} = \sum\limits^{t-1}_{k=0} A_i^{t-k-1}B_iu_{ik} + A_i^tr_{i0}\\
&\forall i, t, ||\overline{r}_{it} - r_{i,goal}||_1 \leq M\cdot (1 - d_{it})\\
&\forall i, t, d_{it} \in \{0,1\}\\
&\forall i\sum\limits_{t = 0}^{T_{max}} d_{it} = 1\\
\nonumber&\forall i, t, -i, \bigvee\limits_{p = 1}^{E_{i,-i}} a^T_{i,-i,p} \overline{r}_{it}> b_{i,-i,0,p} + a_{i,-i,p} \cdot \Delta r'_{-i,t} \\
&\quad\quad\quad\quad \quad\quad\quad+ e_{i,-i,t} -M\sum\limits_{k = 0}^t (d_{ik} + d_{-i,k})\label{C:goalma}\\
&\forall i \forall t, \Sigma_{it} = \sum\limits^{t-1}_{k=0}(A_i + K_iB_i)^{t-k-1}\Sigma_i[(A_i + K_iB_i)^T]^{t-k-1} \label{mp3:noise}\\
&\forall i, t, \Delta r'_{it} =  \sum\limits_{k=0}^{t-1} A_{i}^{t-k-1}B_{i}u_{i,k} + A_{i}^tr_{i0} - r_{i0}\\
&\forall i, t\forall -i,e_{i,-i,t} = \sqrt{a^T_{i,-i,p}(\Sigma_{it} + \Sigma_{-i,t})a_{i,-i,p}}\cdot s_i(t,j)\nonumber\\
&\forall i \forall t \forall n, 0 \leq s_i(n,t) \leq M'
\end{align}
The term $-M\sum\limits_{k = 0}^t (d_{ik} + d_{-i,k})$ in
Constraints~\eqref{C:goalma} means that an agent will not be affected
by other agents who have reached their goal position by time step $t$,
and, conversely, it will not affect the final solution once it reaches its goal position.

%% file: experiments.tex
\section{Experiments}

\begin{figure} [h!]
  \centering 
  \subfigure[]{ 
    \label{fig:subfig:se} %% label for first subfigure 
    \includegraphics[width=0.09\textwidth]{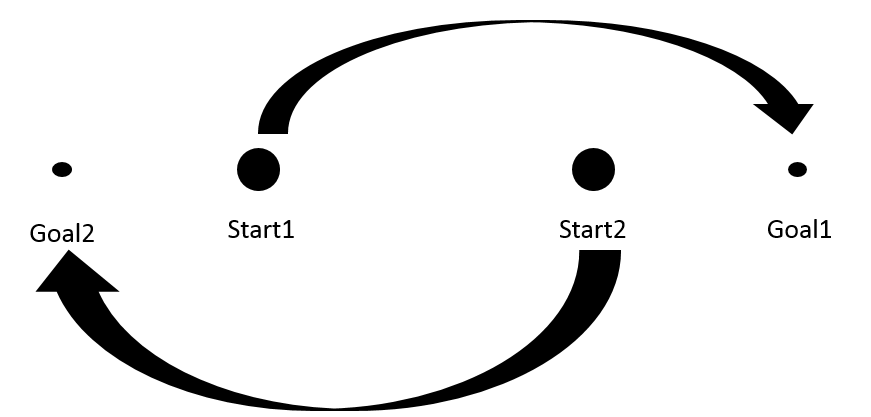}
    } 
  \hspace{1mm} 
  \subfigure[]{ 
    \label{fig:subfig:pm} %% label for second subfigure 
    \includegraphics[width=0.09\textwidth]{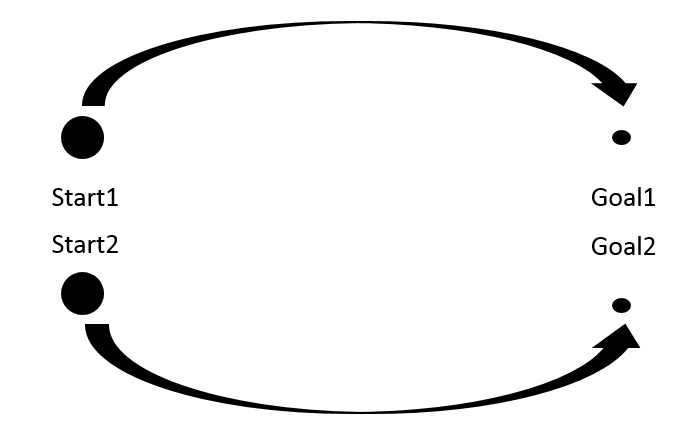}
    
    } 
  \hspace{1mm} 
  \subfigure[]{ 
    \label{fig:subfig:is} %% label for second subfigure 
    \includegraphics[width=0.09\textwidth]{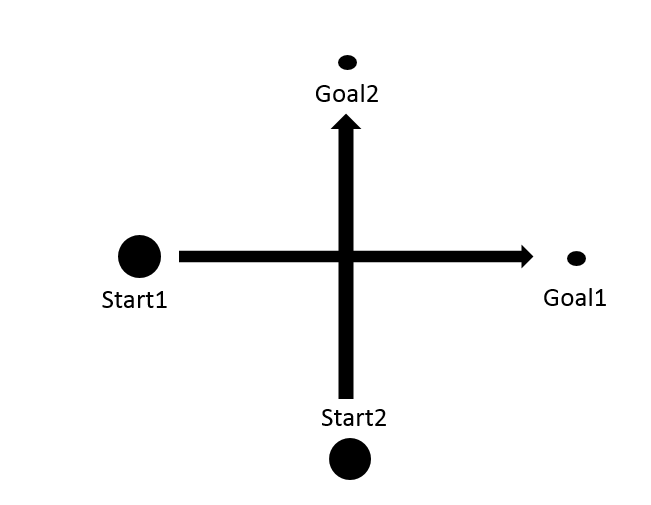}
    } 
    \hspace{1mm} 
  \subfigure[]{ 
    \label{fig:subfig:3is} %% label for second subfigure 
    \includegraphics[width=0.09\textwidth]{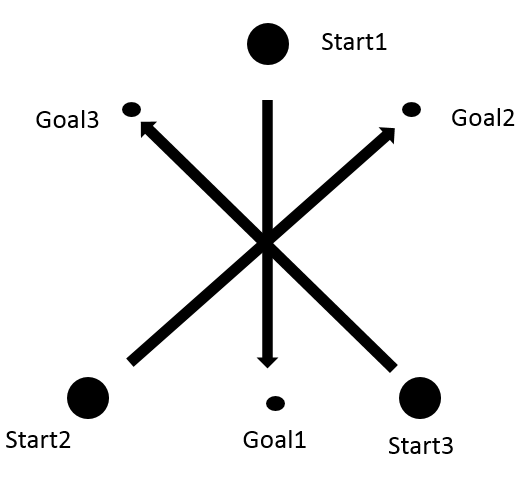}
    } 
  \label{fig:Comparison} %% label for entire figure 
  \caption{Experiment scenarios.} 
  \label{fig:scenario} %% label for entire figure 
\end{figure}

Armed with the techniques for computing both Nash equilibria in path
planning games, as well as a socially optimal solution of the corresponding
``cooperative'' multi-agent planning scenario, we now consider several
case studies to understand the impact of self-interested behavior.
Specifically, we consider the following 2D scenarios:
%Several experiments are applied to test our approach. 
%For the cases with two agents, we consider following three different scenarios. 
\begin{itemize}
\item \textbf{2 agents with opposing goal positions (Figure~\ref{fig:subfig:se}):} the goal position of each
  agent is behind the initial placement of the other.
%, as shown in fig:\ref{fig:subfig:se}. 
In this scenario, the first agent moves from starting coordinate
position $(10,50)$ to goal at position $(95,50)$, and the second agent moves from $(90,50)$ to $(5,10)$.
\item \textbf{2 agents moving in parallel
    (Figure~\ref{fig:subfig:pm}):} the initial and goal positions of
  both agents are near one another.
%are close and two agents trend to move from one side to the other in
%parallel, shown in fig:\ref{fig:subfig:pm}. 
In this scenario, the first agent moves from $(10,70)$ to $(95,70)$ and the second agent moves from $(10,35)$ to $(95,35)$.
\item \textbf{Intersection with 2 agents
    (Figure~\ref{fig:subfig:is}):} 
%in this scenario, 
one agent
  moves from the bottom to the top of the 2D grid, and the other moves
  from left to right.
%, shown in fig:\ref{fig:subfig:is}. 
In this scenario the first agent moves from $(10,50)$ to $(90,50)$ and the second agent moves from $(50,10)$ to $(50,90)$.
\item \textbf{Intersection with 3 agents
    (Figure~\ref{fig:subfig:3is})}: one agent starts at the top of a
  2D grid and moves down, while the other two start at southeast and
  southwest, and move northwest and southeast, respectively. In this
  scenario the first agent moves from $(50,90)$ to $(50,5)$, the
  second agent moves from $(85, 30)$ to $(11, 73)$, and the third agent moves from $(14, 29)$ to $(90, 73)$.
%This is the intersection scenario with 3 players, shown in fig:\ref{fig:subfig:3is}
\end{itemize}
In each experiment, each agent is represented by a square with each
side of length 15 and parallel to either the $x$ or the $y$ axis. 
%(the
%sides of such square parallel either x-axis or y-axis and the
%attachment is the center of the square). 
The control inputs are 2D velocity vectors and the maximum velocity of agents in both $x$ and $y$ direction is 10 (thus, $A = B = I$ in agents' motion dynamic). 
Agents' motion is distorted by a Gaussian distribution with the
covariance matrix $1.9I$. 
For each scenario we consider solutions with and without feedback
control, where the feedback gain for the latter was chosen to be
$K=0.5$.
%Here we consider the cases with and without feedback control scheme
%and set the feedback gain $K$ to $0.5$.
Throughout, we assume that all players are equally concerned about
safety vs.\ efficiency; formally, all players share the same parameter $\lambda$.

The results are shown in
Figures~\ref{fig:ComparisonSide}-\ref{fig:fb_ComparisonIntersection3}.
In each figure, the horizontal axis is the $\lambda$ value
which represents the importance of safety for both agents, where lower
values of $\lambda$ imply that safety is \emph{more} important.
The left plots show the objective value, where lower is better.
The middle plots give the time to goal, where lower is, again, better.
The right plots show safety margin, where again lower is better.
%which we are interested in being
%high (so higher is better).
We present average quantities over all agents; the qualitative
observations are similar if we consider these at individual agent level.

The first observation is that the
difference between socially optimal and equilibrium objective values
appears small ((a) plots in
Figures~\ref{fig:ComparisonSide}-\ref{fig:fb_ComparisonIntersection3}).
%,
%although in several cases we see these diverge somewhat as safety
%becomes more important to the players.
It is therefore tempting to conclude that equilibrium behavior is
similar to socially optimal, but it turns out that this is not the
case: in particular, it turns out that the trade-off between efficiency
and safety made by the agents in equilibrium is very different from
optimal.

Considering next the (b) and (c) columns of the figures, we can
observe that systematically
%The second observation, however, is far more revealing: 
%in almost
%every case, 
performance improves, while safety is often significantly
compromised, in equilibrium as compared to a social optimum.
The difference is particularly dramatic in the first two scenarios, when
the agents are in direct conflict in their quest to reach their
respective goals.
The gap between optimal and equilibrium safety in the other 
scenarios tends to be larger for relatively high values of $\lambda$.
%, but
%remains significant regardless in all but the parallel setting.

Another general observation we can make is that often the solutions
with a feedback controller are closer to optimal, particularly from
the perspective of safety.
The exceptions involve the intersection scenarios, where the gap is
larger for higher values of $\lambda$ in the feedback controller
solution than with the open-loop controller.
However, even in these scenarios, the feedback controller yields
solutions closer to socially optimal for most values of $\lambda$.
This is not surprising: since all agents are concerned about safety,
they are more able to dynamically adjust to avoid collisions when some
feedback about state is available.

To understand why safety is systematically compromised, consider a
single agent's incentive.
Even though an agent is interested in reaching the goal safely, it
does not account for the fact that being involved in a crash
\emph{also crashes the other agent}.
Thus, in equilibrium safety is compromised relative to social optimum,
as agents fail to capture the externalities associated with crashes.

\begin{figure*} [htb]
  \centering 
  \subfigure[]{ 
    \includegraphics[width=0.31\textwidth, height = 22 mm]{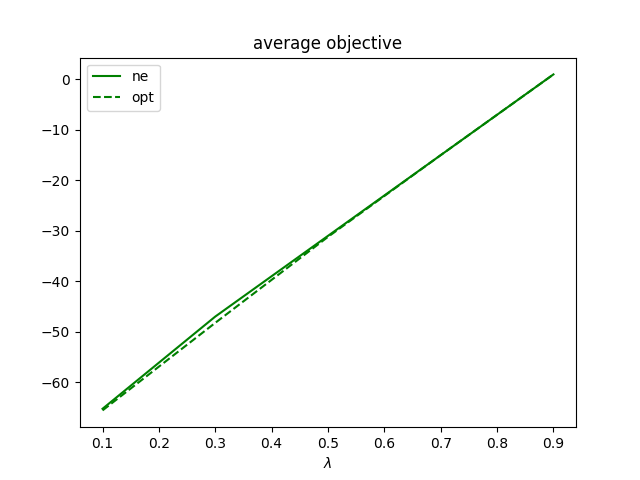}} 
  \subfigure[]{ 
    \includegraphics[width=0.31\textwidth, height = 22 mm]{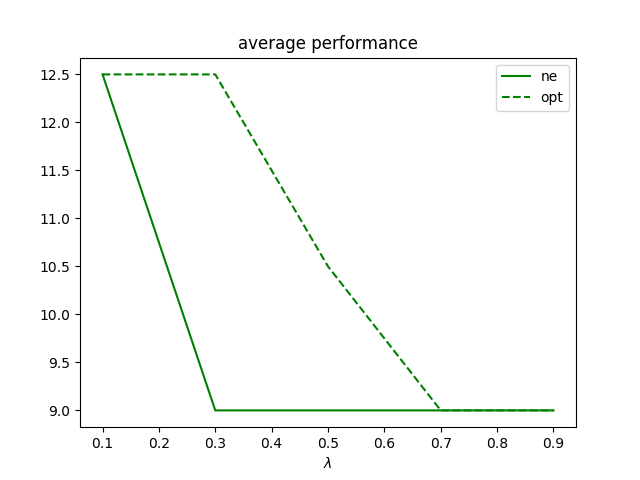}} 
  \subfigure[]{ 
    \includegraphics[width=0.31\textwidth, height = 22 mm]{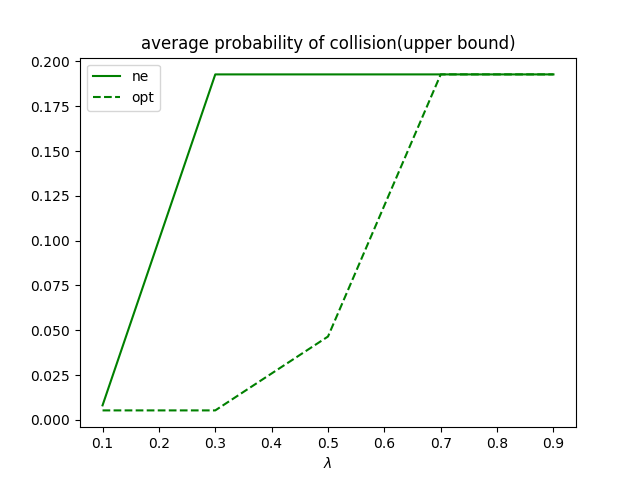}} 
  \caption{Opposing goal positions without the feedback gain($K=0$).} 
  \label{fig:ComparisonSide} %% label for entire figure 

  \centering 
  \subfigure[]{ 
    \includegraphics[width=0.31\textwidth, height = 22 mm]{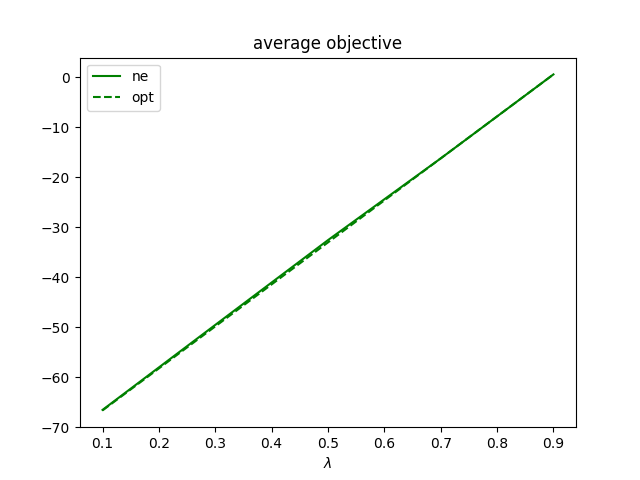}} 
  \subfigure[]{ 
    \includegraphics[width=0.31\textwidth, height = 22 mm]{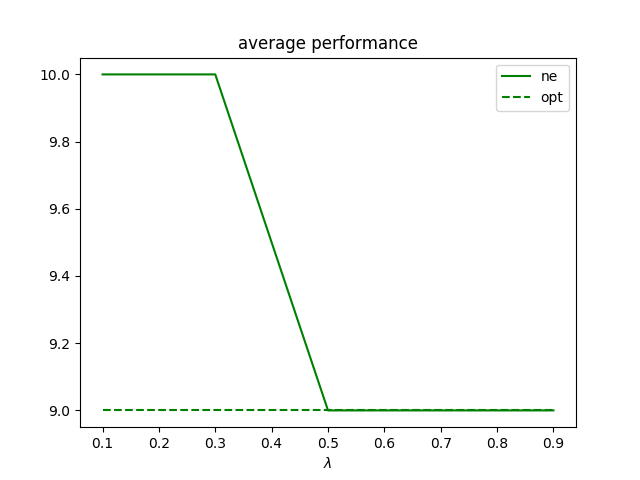}} 
  \subfigure[]{ 
    \includegraphics[width=0.31\textwidth, height = 22 mm]{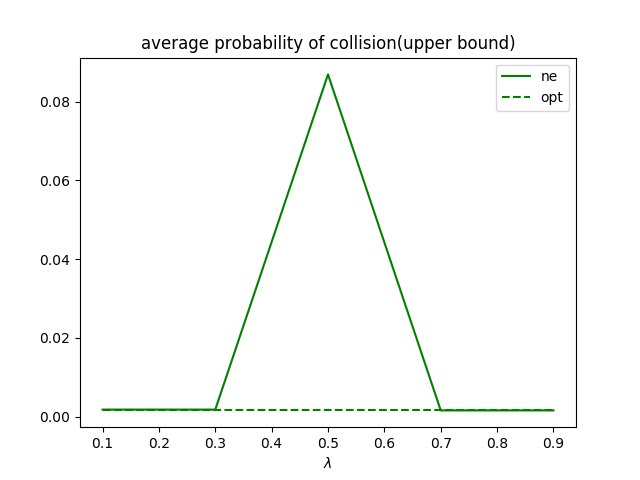}} 
  \caption{Opposing goal positions with the feedback gain($K = 0.5$).} 
  \label{fig:fb_ComparisonSide} %% label for entire figure 

  \end{figure*}
  \clearpage   
  \setcounter{figure}{5}
\begin{figure*}[htb]\ContinuedFloat
  \centering 
  \subfigure[]{ 
    \includegraphics[width=0.31\textwidth, height = 22 mm]{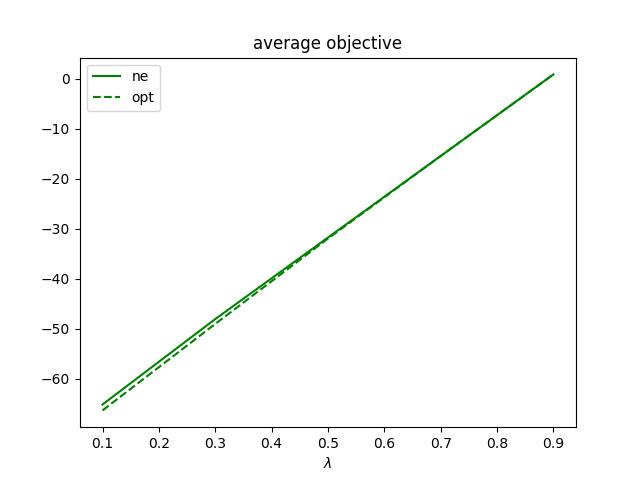}} \subfigure[]{
    \includegraphics[width=0.31\textwidth, height = 22 mm]{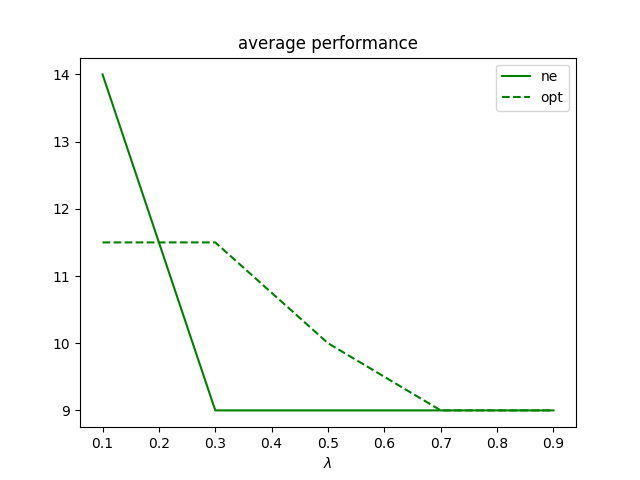}}  
  \subfigure[]{ 
    \includegraphics[width=0.31\textwidth, height = 22 mm]{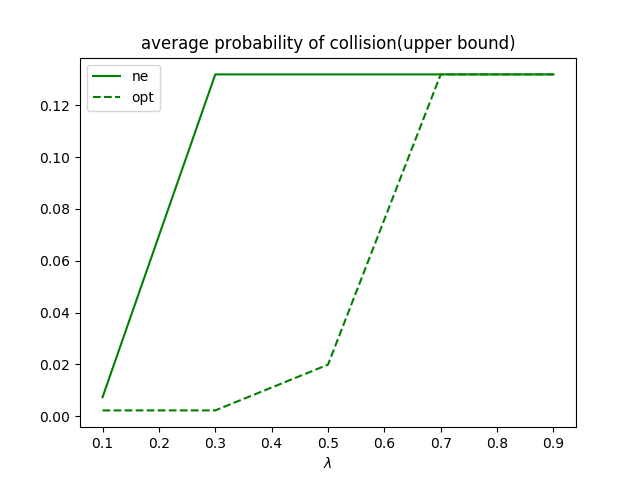}} 
  \caption{Moving in parallel without the feedback gain($K=0$).} 
  \label{fig:ComparisonParallel} %% label for entire figure 
\centering 
  \subfigure[]{ 
    \includegraphics[width=0.31\textwidth, height = 22 mm]{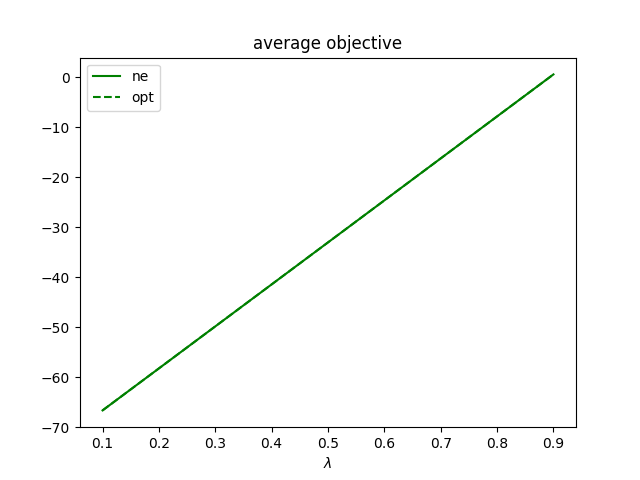}} 
  \subfigure[]{ 
    \includegraphics[width=0.31\textwidth, height = 22 mm]{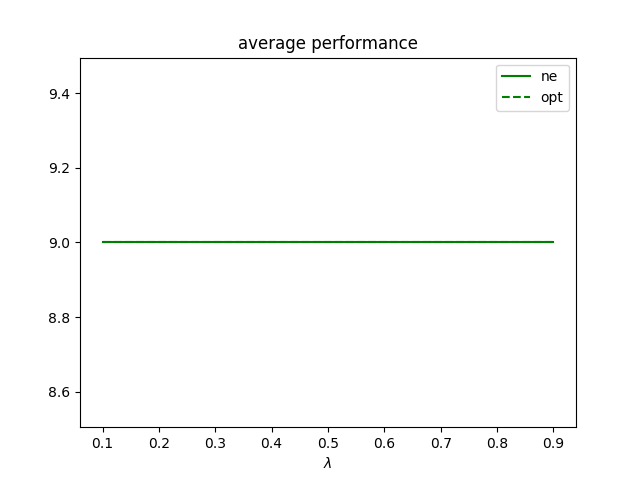}} 
  \subfigure[]{ 
    \includegraphics[width=0.31\textwidth, height = 22 mm]{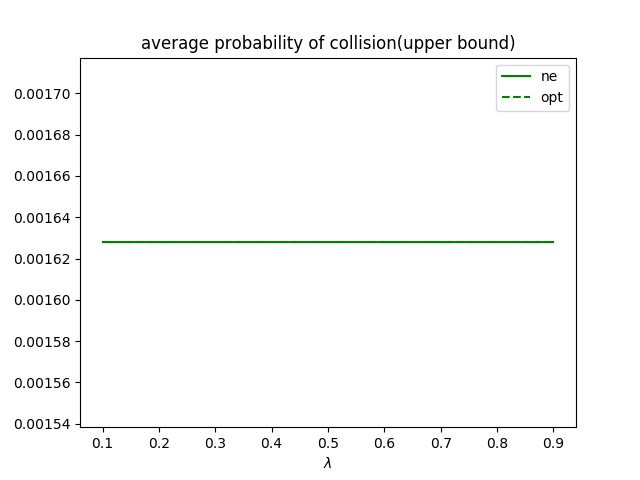}} 
  \caption{Moving in parallel with the feedback gain($K = 0.5$).} 
  \label{fig:fb_ComparisonParallel} %% label for entire figure 

  \centering 
  \subfigure[]{  
    \includegraphics[width=0.31\textwidth, height = 22 mm]{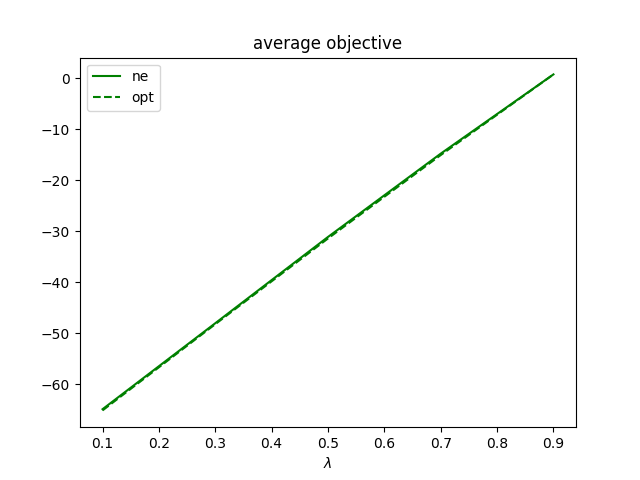}} 
 % \hspace{1mm} 
  \subfigure[]{  
    \includegraphics[width=0.31\textwidth, height = 22 mm]{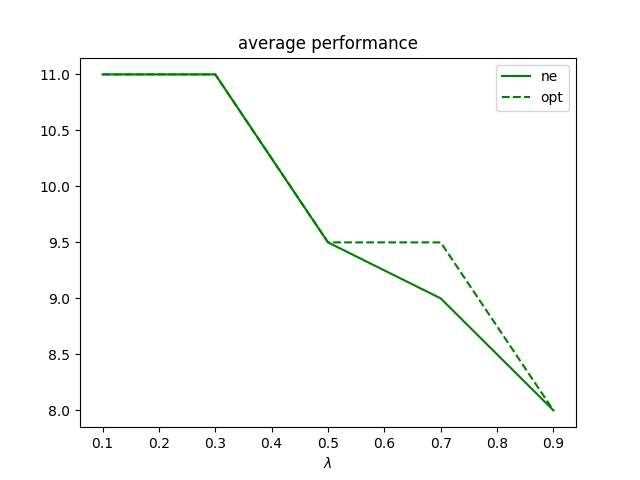}} 
  \subfigure[]{ 
    \includegraphics[width=0.31\textwidth, height = 22 mm]{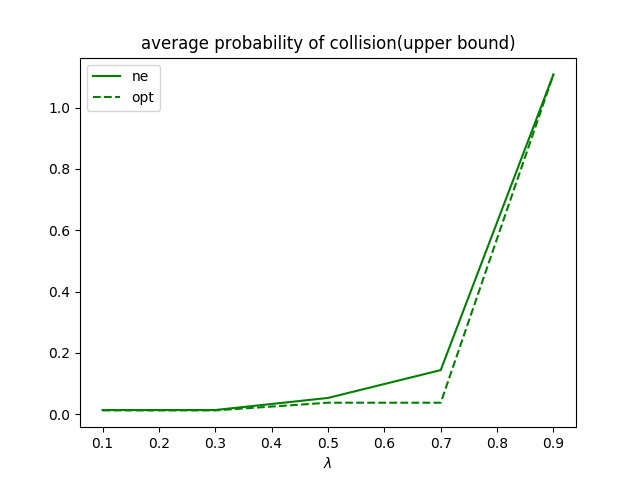}} 
  \caption{Intersection without the feedback gain($K=0$, 2 agents).} 
 \label{fig:ComparisonIntersection2} 
   \centering 
  \subfigure[]{ 
    \includegraphics[width=0.31\textwidth, height = 22 mm]{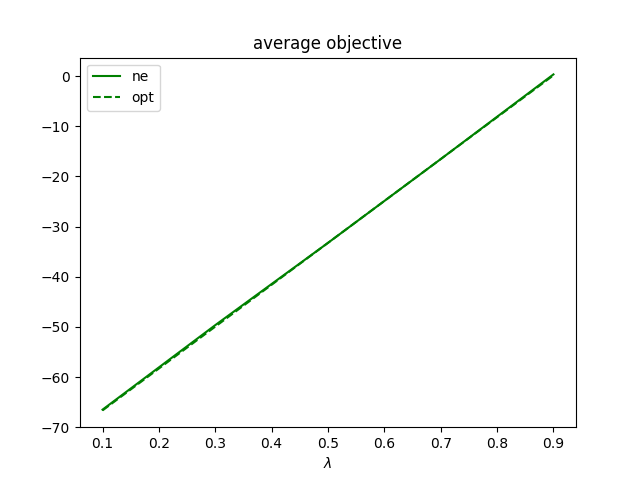}} 
  %\hspace{1mm} 
  \subfigure[]{ 
    \includegraphics[width=0.31\textwidth, height = 22 mm]{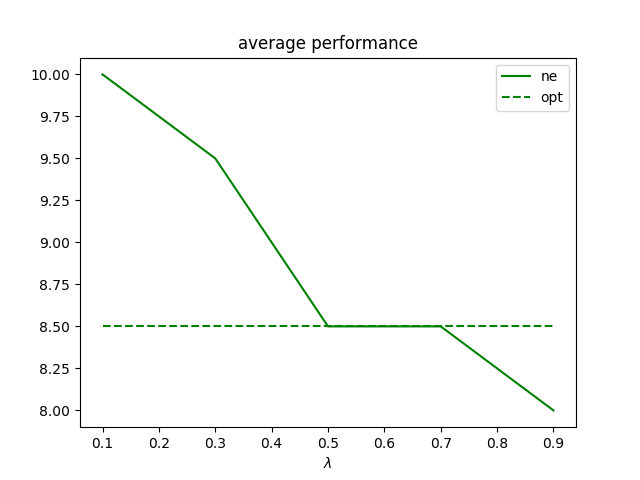}} 
  %\hspace{1mm} 
  \subfigure[]{ 
    \includegraphics[width=0.31\textwidth, height = 22 mm]{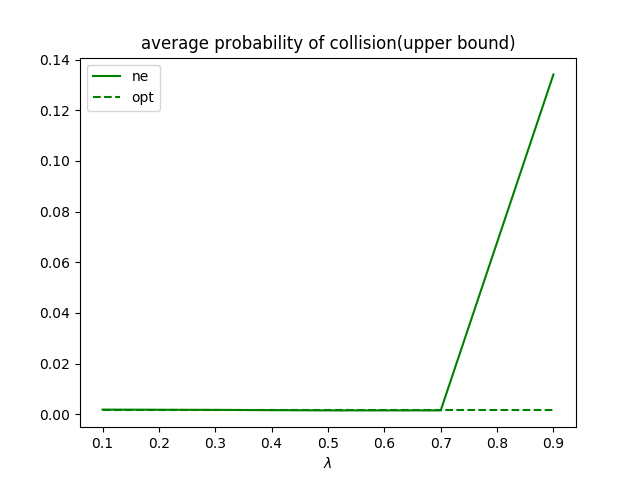}} 
  \caption{Intersection with the feedback gain($K = 0.5$, 2 agents).} 
 \label{fig:fb_ComparisonIntersection2} 
\centering 
  \subfigure[]{ 
    \includegraphics[width=0.31\textwidth, height = 22 mm]{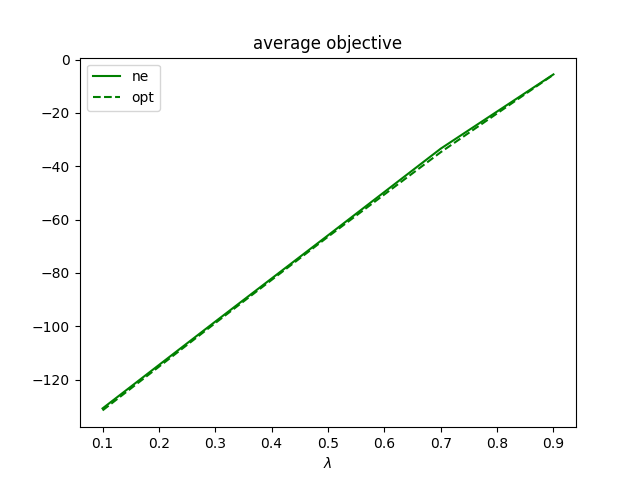}} 
  %\hspace{1mm} 
  \subfigure[]{ 
    \includegraphics[width=0.31\textwidth, height = 22 mm]{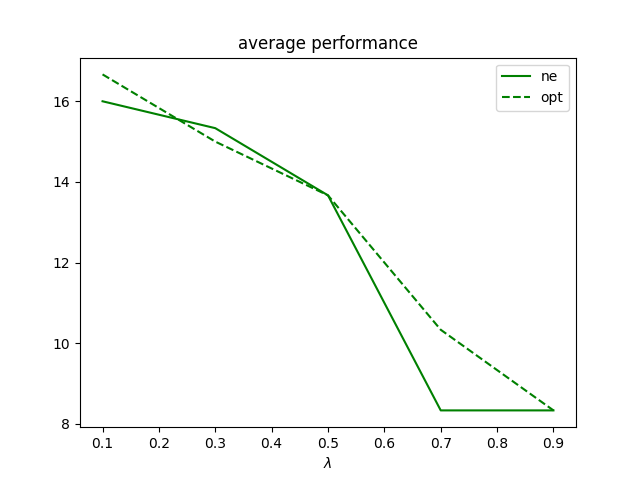}} 
  %\hspace{1mm} 
  \subfigure[]{ 
    \includegraphics[width=0.31\textwidth, height = 22 mm]{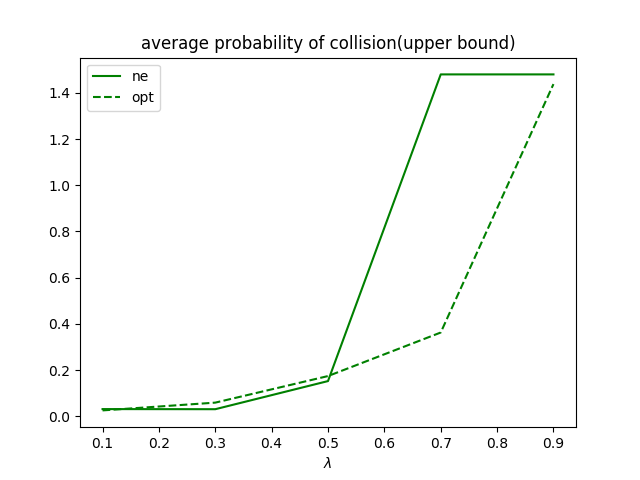}} 
    \caption{Intersection  without the feedback gain($K=0$, 3 players).} 
  \label{fig:ComparisonIntersection3} %% label for entire figure 
\centering 
  \subfigure[]{ 
    \includegraphics[width=0.31\textwidth, height = 22 mm]{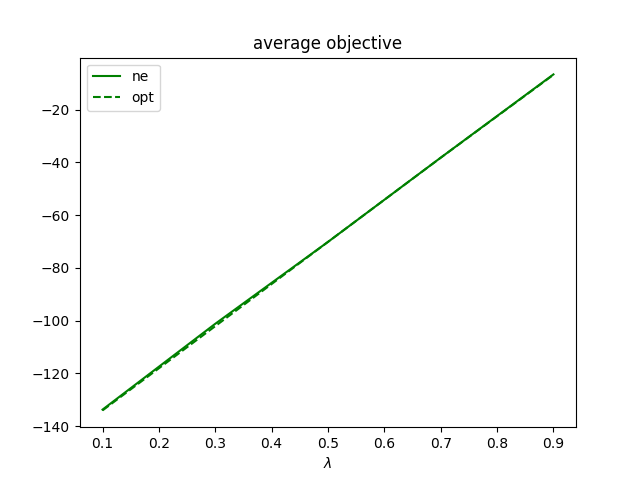}} 
  %\hspace{1mm} 
  \subfigure[]{ 
    \includegraphics[width=0.31\textwidth, height = 22 mm]{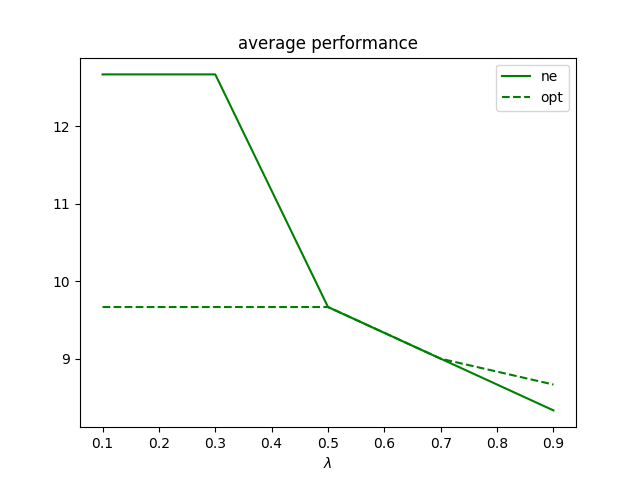}} 
  %\hspace{1mm} 
  \subfigure[]{ 
    \includegraphics[width=0.31\textwidth, height = 22 mm]{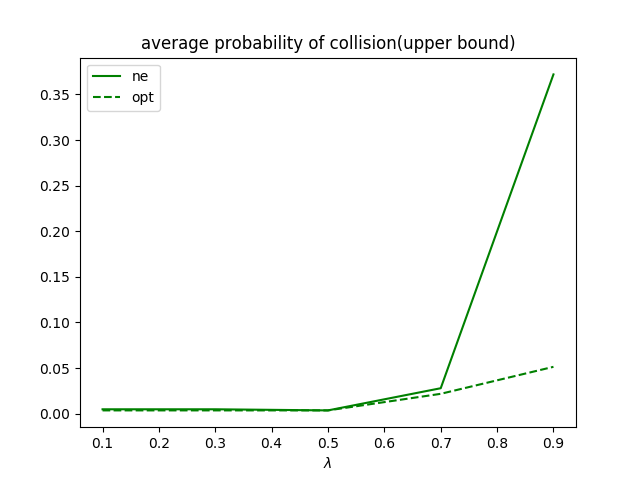}} 
    \caption{Intersection with the feedback gain($K = 0.5$, 3 players).} 
  \label{fig:fb_ComparisonIntersection3} %% label for entire figure 

\end{figure*}

%% file: main.bbl
\begin{thebibliography}{24}
\providecommand{\natexlab}[1]{#1}
\providecommand{\url}[1]{\texttt{#1}}
\providecommand{\urlprefix}{}

\bibitem[{Arkin(1989)}]{Arkin1989}
Arkin, R.C.: {Motor Schema-Based Mobile Robot Navigation} (1989)

\bibitem[{{Auat Cheein} and Carelli(2013)}]{AuatCheein2013}
{Auat Cheein}, F.A., Carelli, R.: {Agricultural robotics: Unmanned robotic
  service units in agricultural tasks}.
\newblock IEEE Industrial Electronics Magazine 7(3), 48--58 (2013)

\bibitem[{Beckmann et~al.(1956)Beckmann, McGuire, and Winsten}]{beckmann1956}
Beckmann, M., McGuire, C., Winsten, C.: {Studies in the Economics of
  Transportation}  (1956)

\bibitem[{Bertsekas(1999)}]{Bertsekas1999}
Bertsekas, D.: {Nonlinear programming} (1999)

\bibitem[{Blackmore et~al.(2006)Blackmore, Li, and
  Williams}]{blackmore2006probabilistic}
Blackmore, L., Li, H., Williams, B.: A probabilistic approach to optimal robust
  path planning with obstacles.
\newblock In: American Control Conference (2006)

\bibitem[{Braess(1968)}]{Braess1968}
Braess, D.: {{\"{U}}ber ein Paradoxon aus der Verkehrsplanung}.
\newblock Unternehmensforschung Operations Research - Recherche
  Op{\'{e}}rationnelle 12(1), 258--268 (dec 1968)

\bibitem[{Chen et~al.(2014{\natexlab{a}})Chen, Zhou, and
  Tomlin}]{chen2014multiplayer}
Chen, M., Zhou, Z., Tomlin, C.J.: Multiplayer reach-avoid games via low
  dimensional solutions and maximum matching.
\newblock In: American Control Conferenc. pp. 1444--1449 (2014{\natexlab{a}})

\bibitem[{Chen et~al.(2014{\natexlab{b}})Chen, Zhou, and Tomlin}]{chen2014path}
Chen, M., Zhou, Z., Tomlin, C.J.: A path defense approach to the multiplayer
  reach-avoid game.
\newblock In: Annual Conference on Decision and Control. pp. 2420--2426
  (2014{\natexlab{b}})

\bibitem[{Craighead et~al.(2007)Craighead, Murphy, Burke, and
  Goldiez}]{Craighead2007}
Craighead, J., Murphy, R., Burke, J., Goldiez, B.: {A survey of commercial {\&}
  open source unmanned vehicle simulators}.
\newblock In: Proceedings - IEEE International Conference on Robotics and
  Automation. pp. 852--857 (2007)

\bibitem[{DeSouza and Kak(2002)}]{DeSouza2002}
DeSouza, G.N., Kak, A.C.: {Vision for mobile robot navigation: A survey}.
\newblock IEEE Transactions on Pattern Analysis and Machine Intelligence 24(2),
  237--267 (2002)

\bibitem[{Elfes(1989)}]{Elfes1989}
Elfes, A.: {Using occupancy grids for mobile robot perception and navigation}.
\newblock Computer 22(6), 46--57 (1989)

\bibitem[{Geibel and Wysotzki(2005)}]{geibel2005risk}
Geibel, P., Wysotzki, F.: Risk-sensitive reinforcement learning applied to
  control under constraints.
\newblock J. Artif. Intell. Res. 24, 81--108 (2005)

\bibitem[{Jonsson and Rovatsos(2011)}]{jonsson2011scaling}
Jonsson, A., Rovatsos, M.: Scaling up multiagent planning: A best-response
  approach.
\newblock In: ICAPS (2011)

\bibitem[{Jord{\'a}n et~al.(2017)Jord{\'a}n, Torreno, de~Weerdt, and
  Onaindia}]{jordan2017better}
Jord{\'a}n, J., Torreno, A., de~Weerdt, M., Onaindia, E.: A better-response
  strategy for self-interested planning agents.
\newblock Applied Intelligence pp. 1--21 (2017)

\bibitem[{LaValle(2000)}]{lavalle2000robot}
LaValle, S.M.: Robot motion planning: A game-theoretic foundation.
\newblock Algorithmica 26(3-4), 430--465 (2000)

\bibitem[{Mahony and Kumar(2012)}]{Mahony2012}
Mahony, R., Kumar, V.: {Aerial robotics and the quadrotor}.
\newblock IEEE Robotics and Automation Magazine 19(3), 19 (2012)

\bibitem[{Oldewurtel et~al.(2008)Oldewurtel, Jones, and
  Morari}]{oldewurtel2008tractable}
Oldewurtel, F., Jones, C.N., Morari, M.: A tractable approximation of chance
  constrained stochastic mpc based on affine disturbance feedback.
\newblock In: IEEE Conference on Decision and Control. pp. 4731--4736 (2008)

\bibitem[{Pigou(1932)}]{Pigou1932}
Pigou, A.: {The economics of welfare, 1920}.
\newblock McMillan{\&}Co., London  (1932)

\bibitem[{Rosenschein and Zlotkin(1994)}]{rosenschein1994rules}
Rosenschein, J.S., Zlotkin, G.: Rules of encounter: designing conventions for
  automated negotiation among computers.
\newblock MIT press (1994)

\bibitem[{Roughgarden and Tardos(2000)}]{Roughgarden2000}
Roughgarden, T., Tardos, E.: {How bad is selfish routing?}
\newblock Proceedings 41st Annual Symposium on Foundations of Computer Science
  49(2), 1--26 (2000)

\bibitem[{Schmeidler(1973)}]{Schmeidler1973}
Schmeidler, D.: {Equilibrium points of nonatomic games}.
\newblock Journal of Statistical Physics 7(4), 295--300 (1973)

\bibitem[{Shen et~al.(2008)Shen, Chen, Cruz, and Blasch}]{shen2008game}
Shen, D., Chen, G., Cruz, J.B., Blasch, E.: A game theoretic data fusion aided
  path planning approach for cooperative {UAV ISR}.
\newblock In: Aerospace Conference. pp. 1--9 (2008)

\bibitem[{Stone and Veloso(2000)}]{stone2000multiagent}
Stone, P., Veloso, M.: Multiagent systems: A survey from a machine learning
  perspective.
\newblock Autonomous Robots 8(3), 345--383 (2000)

\bibitem[{Wardrop(1952)}]{wardrop1952road}
Wardrop, J.G.: Road paper. some theoretical aspects of road traffic research.
\newblock Proceedings of the institution of civil engineers 1(3), 325--362
  (1952)

\end{thebibliography}
